\documentclass[apjl,twocolumn]{emulateapj}
\usepackage{natbib}
\newcommand{\lsim}{\raisebox{-0.13cm}{~\shortstack{$<$ \\[-0.07cm] $\sim$}}~}
\newcommand{\gsim}{\raisebox{-0.13cm}{~\shortstack{$>$ \\[-0.07cm] $\sim$}}~}
\newcommand{\ffm}{4.5 \, \rm  \mu m}
\newcommand{\tfm}{24 \, \rm  \mu m}

\shorttitle{The Nature of Extremely Red $H-[4.5]>4$ Galaxies}
\shortauthors{Caputi et al.}

\begin{document}

\title{The Nature of Extremely Red $H-[4.5]>4$ Galaxies \\
    revealed with SEDS and CANDELS}


\author{K. I. Caputi\altaffilmark{1,2},
J. S. Dunlop\altaffilmark{1},
R. J. McLure\altaffilmark{1},
J.-S. Huang\altaffilmark{3},
G. G. Fazio\altaffilmark{3},
M. L. N. Ashby\altaffilmark{3},
M. Castellano\altaffilmark{4},
A. Fontana\altaffilmark{4},
M. Cirasuolo\altaffilmark{1,5},
O. Almaini\altaffilmark{6},
E. F. Bell\altaffilmark{7},
M. Dickinson\altaffilmark{8}, 
J. L. Donley\altaffilmark{9}, 
S. M. Faber\altaffilmark{10}, 
H. C. Ferguson\altaffilmark{9},
M. Giavalisco\altaffilmark{11},
N. A. Grogin \altaffilmark{9},
D. D. Kocevski\altaffilmark{10},
A. M. Koekemoer\altaffilmark{9},
D. C. Koo\altaffilmark{10},
K. Lai \altaffilmark{10},
J. A. Newman\altaffilmark{12},
R. S. Somerville\altaffilmark{13}
}



\altaffiltext{1}{SUPA, Institute for Astronomy, The University of Edinburgh, Royal Observatory, Edinburgh, EH9 3HJ, UK}
\altaffiltext{2}{Present address: Kapteyn Astronomical Institute, University of Groningen, P.O. Box 800, 9700 AV Groningen,
The Netherlands. Email: karina@astro.rug.nl}
\altaffiltext{3}{Harvard-Smithsonian Center for Astrophysics, 60 Garden Street, Cambridge, MA 02138, USA}
\altaffiltext{4}{INAF - Osservatorio Astronomico di Roma, Via Frascati 33, I--00040, Monteporzio, Italy}
\altaffiltext{5}{Astronomy Technology Center, Royal Observatory, Edinburgh EH9 3HJ, UK}
\altaffiltext{6}{School of Physics and Astronomy, University of Nottingham, University Park, Nottingham NG7 2RD, UK}
\altaffiltext{7}{Department of Astronomy, University of Michigan, 500 Church St., Ann Arbor, MI 48109, USA}
\altaffiltext{8}{National Optical Astronomy Observatory, 950 N. Cherry Ave., Tucson AZ 85719 USA}
\altaffiltext{9}{Space Telescope Science Institute, 3700 San Martin Drive, Baltimore, MD 21218, USA}
\altaffiltext{10}{University of California Observatories/Lick Observatory, University of California, Santa Cruz, CA 95064, USA}
\altaffiltext{11}{Department of Astronomy, University of Massachusetts, Amherst, MA 01003, USA}
\altaffiltext{12}{University of Pittsburgh, Department of Physics and Astronomy and PITT-PAC, Pittsburgh, PA, USA}
\altaffiltext{13}{Physics and Astronomy Department, Rutgers University, Piscataway, NJ 08854, USA}

\begin{abstract}
We have analysed a sample of 25 extremely red $H-[4.5]>4$ galaxies, selected using  $\ffm$ 
data from the {\em Spitzer} SEDS survey and deep $H$-band data from the {\em Hubble Space Telescope (HST)} CANDELS survey, over $\sim$180 arcmin$^2$ of the UKIDSS Ultra Deep Survey (UDS) field.  Our aim is to investigate the nature of this rare population of mid-infrared (mid-IR) sources that display such extreme near-to-mid-IR colours.  Using up to 17-band photometry  ($U$ through 8.0~$\rm \mu m$), we have studied in detail their spectral energy distributions, including possible degeneracies in the photometric redshift/internal extinction  ($z_{\rm phot}-A_V$) plane. Our sample appears to include sources of very different nature. Between 45\% and 75\% of them are dust-obscured, massive galaxies at $3<z_{\rm phot}<5$. All of the $\tfm$-detected sources in our sample are in this category. Two of these have $S(24 \, \rm \mu m)>300 \, \rm \mu Jy$, which at $3<z_{\rm phot}<5$ suggests that they probably host a dust-obscured active galactic nucleus (AGN). Our sample also contains four highly obscured ($A_V>5$) sources at $z_{\rm phot}<1$. Finally, we analyse in detail  two $z_{\rm phot}\sim6$ galaxy candidates, and discuss their plausibility and implications. Overall, our red galaxy sample contains the tip of the iceberg of a larger population of $z>3$ galaxies to be discovered with the future {\em James Webb Space Telescope}.
\end{abstract}


\section{Introduction}
\label{sec-intro}

The discovery of new galaxy populations at high redshifts ($z>3$) is one of the most exciting goals of deep galaxy surveys. Currently, rapid progress is being made in this field, thanks to several deep multi-wavelength surveys that are being conducted with world-class ground-based and space telescopes.  These new galaxy surveys can potentially make a breakthrough in the field by: i) discovering galaxies beyond the maximum known galaxy redshift limit; ii) revealing a previously unrecognised galaxy population whose properties are significantly different to those of other known galaxies at similar redshifts.  The latter is crucial to achieve an unbiased view of how galaxy evolution proceeded over the first few billion years of cosmic time.

The selection of galaxies with very red near-IR/optical colours has proven very useful to identify the high-$z$ tail of different galaxy populations. This is the underlying selection technique of what are traditionally called extremely red galaxies, i.e. galaxies with $I-K>4$ or $R-K>5$ Vega mag \citep[e.g.][]{roc03,bro05}, whose redshift distribution is confined to $z\gsim 1$ \citep{cap04}. Another kind of near-IR colour cut ($J-K>2.3$ Vega mag) has been applied to select `distant red galaxies' at $z\sim 2-3$ \citep{fra03}.  The basis of these colour techniques lies in bracketing the galaxy 4000~$\rm \AA$ break, which is shifted into the observed near-IR bands at $z \gsim 1.5$, although it is known that similar colours can also be displayed by galaxies with highly reddened spectral energy distributions (SEDs) \citep[see e.g.][]{sma02,pap06,fon09}.  Optically obscured sources selected at mid-IR wavelengths have also been identified as $z \gsim 2$ galaxies \citep{hou05,dey08}.

At $z>3$, the 4000 $\rm \AA$ break is shifted to wavelengths $\lambda > 1.6 \, \rm \mu m$, implying that the potentially oldest and/or dustier galaxies at those redshifts could be missed even by deep near-IR surveys. This is indeed the case for the sub-millimetre source GN10 \citep{wan07}, which has no near-IR counterpart down to a magnitude of $K_s \approx 27$ (AB), but is detected at mid-IR wavelengths $\lambda \geq  3.6 \, \rm \mu m$. Very recently, Huang et al.~(2011) reported the existence of four galaxies of a similar kind, selected with the {\em Spitzer Space Telescope} Infrared Array Camera \citep[IRAC;][]{faz04}. These galaxies have colours $H-[3.6]>4.5$ (AB),  and their SED fitting suggests that they lie at $z\sim4-6$.

In this work we have undertaken a combined analysis of the {\em Spitzer} Extended Deep Survey (SEDS; Ashby et al.~2012, in preparation) with the {\em HST} Cosmic Assembly Near-infrared Deep Extragalactic Legacy Survey \citep[CANDELS; ][]{gro11,koe11}, to search for extremely red $H-[4.5]>4$ (AB) galaxies over $\sim$180 arcmin$^2$ of the UKIDSS UDS field. Our aim is to study the variety in composition of this colour-selected galaxy population, and investigate the existence of  high-$z$ galaxy candidates. All magnitudes and colours quoted in this paper are total and refer to the AB system \citep{oke83}, unless otherwise stated.  We adopt a cosmology with  $\rm H_0=70 \,{\rm km \, s^{-1} Mpc^{-1}}$, $\rm \Omega_M=0.3$ and $\rm \Omega_\Lambda=0.7$. All stellar masses refer to a Salpeter (1955) initial mass function (IMF) over stellar masses $(0.1-100) \, \rm M_\odot$.

\section{Multi-wavelength datasets and sample selection}
\label{sec-sample}

The Warm {\em Spitzer}/IRAC SEDS survey has mapped five different fields of the sky at 3.6 and $\ffm$, over a total area of 0.90 deg$^2$, with an integration time of 12 hours/pointing. In this work we analyse the $\sim$180 arcmin$^2$ of the  SEDS UDS field that overlap the  {\em HST} CANDELS survey \citep{gro11,koe11}, which comprises observations with the Wide Field Camera 3 (WFC3) F125W ($J$) and F160W ($H$) filters, and the Advanced Camera for Surveys (ACS) F606W and F814W filters. 

The UDS field also benefits from multi-wavelength ancillary data, including UV and optical data from Subaru and the Canada-France Hawaii Telescope (CFHT), near-IR data from the UK Infrared Telescope (UKIRT), and shallower {\em Spitzer} data from the SpUDS Legacy Survey (P.I. Dunlop). We have used all these datasets to constrain the SEDs of our $H-[4.5]>4$ selected galaxies, whenever possible.

We have extracted a source catalogue from the IRAC SEDS/UDS $\ffm$  image using the software SEXTRACTOR \citep{ba96} following a similar procedure as that described in Caputi et al.~(2011). As in this previous work, we have performed simulations and determined that the 80\% completeness limit of our IRAC SEDS/UDS $\ffm$ catalogue corresponds to a magnitude [4.5]=23.9.  This limit is 1.5 mag deeper than the 80\% completeness limit of the SpUDS $\ffm$ catalogue \citep{cap11}.  We measured  3.6 $\rm \mu m$ photometry for the IRAC SEDS $\ffm$ sources using SEXTRACTOR in dual-image mode. All our IRAC magnitudes have been obtained from 4~arcsec-diameter aperture magnitudes plus aperture corrections ($\approx -0.30$ mag). 

Independently, we have extracted a source catalogue from the CANDELS WFC3 $H$-band map, and measured photometry on the ACS F606W and F814W, and WFC3 F125W maps using SEXTRACTOR in dual-image mode. The depth of the CANDELS/UDS $H$-band map is of $H\approx 27.1$ mag \citep[5$\sigma$; ][]{gro11}. All our {\em HST} magnitudes are obtained from 2~arcsec-diameter aperture magnitudes with very small ($\approx-0.05$ mag) aperture corrections.

\subsection{$H-[4.5]>4$ galaxies in the SEDS/CANDELS UDS field}

We have searched for counterparts to our IRAC SEDS [4.5]$\leq$23.9 sources in the CANDELS  $H$-band-selected catalogue, within a matching radius $r=1.5 \, \rm arcsec$. We restricted our analysis to the IRAC SEDS sources within the $\ffm$ catalogue 80\% completeness limit simply to have sufficient colour range to identify extremely red sources. For the identifications, we have used the entire $H$-band catalogue with no magnitude cut.

After excluding sources with contaminated photometry in the proximity of bright stars, we have identified 25 IRAC SEDS sources with  single CANDELS counterparts within $r=1.5 \, \rm arcsec$ radius and colours $H-[4.5]>4$. Note that the excellent SEDS/CANDELS relative astrometry means that, in practice, our sources are identified within a radius $r \lsim 0.5 \, \rm arcsec$, but inspecting a $r=1.5 \, \rm arcsec$ circular area has allowed us to eliminate cases of blended photometry.

We have also visually inspected our IRAC SEDS  sources that do not have a counterpart in the CANDELS maps. There are 18 $\ffm$ sources (also detected at $3.6 \, \rm \mu m$) which are truly near-IR `dropouts' at the depth of the CANDELS images. We note that, given the depths of the SEDS $\ffm$ and CANDELS maps, some of these near-IR dropouts might not strictly have $H-[4.5]>4$ colours. Nonetheless, we will include them all in the discussion presented in Section \S\ref{sec-col}.  

We note that the extremely red near-to-mid-IR colours ($H-[4.5]>4$) of our galaxies cannot be produced by M, L or T-type dwarf stars \citep{cus05,ray09}, which instead can display extremely red optical-to-near-IR colours \citep[e.g.][]{ste07}.

\section{Analysis of the CANDELS-detected $H-[4.5]>4$ galaxies}
\label{sec-reddet}

\begin{figure}
\plotone{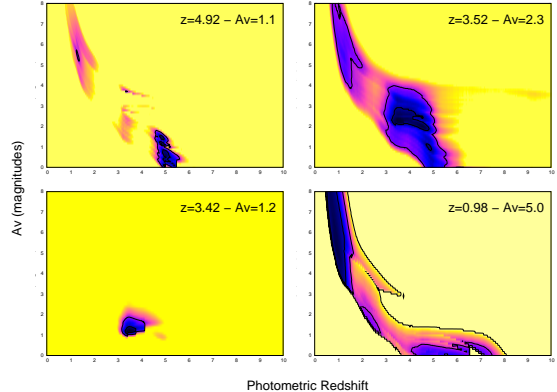}
\caption{Examples of our SED fitting results for $H~-~[4.5]~>~4$ galaxies. Each plot shows 1,2 and 3$\sigma$ confidence levels on the $z_{\rm phot}-A_V$ plane, around the solution producing the minimum $\chi^2$ value (top-right corner). \label{fig_sedfit}}
\end{figure}

We have compiled up to 17-band photometry ($U$ through $8.0 \, \rm \mu m$) for our 25  CANDELS-detected $H-[4.5]>4$ galaxies, in order to model their SEDs, and constrain their redshifts and derived parameters. The shallow SpUDS 5.8 and $8.0 \, \rm \mu m$ data were included when available,  which in some cases helped mitigate degeneracies in parameter space. For the SED modelling we have used a customised $\chi^2$-minimisation SED-fitting code with a library that included synthetic stellar templates from the 2007 version of  the Bruzual \& Charlot models \citep{bch03,bru07}, with solar metallicity, and two empirical templates of a type-2 QSO and a QSO/star-forming composite galaxy from the library by Polletta et al.~(2006). To account for internal extinction, we have convolved the stellar templates with the Calzetti et al.~(2000) reddening law, allowing for $0.0 \leq A_V \leq 8.0$ with a step of 0.1.  In the case of non-detections, we have not allowed for any modelled SED that produced a flux above the $\sim70\%$ completeness detection limit in the corresponding waveband. 

An assessment of the photometric redshift quality for galaxies in our parent $\ffm$ sample has been presented in Caputi et al.~(2011): the overall dispersion is 5\%, and $\sim 12\%$ for the $z>2$ sources. An important difference  with respect to this paper is that we now include the deep CANDELS data, which makes possible the identification and analysis of the  $H~-~[4.5]~>~4$ sources.
 
We show examples of our SED fitting results in Fig.~\ref{fig_sedfit}.  Not surprisingly, in most cases the best-fitting solutions are degenerate in $z_{\rm phot}-A_V$ space, i.e. there are different $z_{\rm phot}-A_V$ combinations within 3$\sigma$ confidence of the minimum  $\chi^2$ value.  In spite of this,  the best-fitting solutions can be well identified in 17 out of our 25 $H-[4.5]>4$ galaxies. For the remaining eight,  the resulting SEDs are `unconstrained', i.e. the fitting degeneracies are so strong that the $z_{\rm phot}$ probability density is virtually flat, so we cannot derive any clear redshift estimates.   We provide the coordinates and photometry of all our sources in Table~1.  The 17 sources with constrained SEDs can be classified in the following groups.

\subsection{Galaxies at $3<z_{\rm phot}<5$}

This is the largest group within our sample: eleven out of 25 sources appear to be at $3<z_{\rm phot}<5$, with extinction values in the range $1.1 \leq A_V \leq 4.2$ (see examples in Fig.~\ref{fig_sedfit}).  

We found that six out of the eleven $3<z_{\rm phot}<5$ galaxies are detected in the SpUDS $\tfm$
images. For all these galaxies the best-fitting SED models are obtained with stellar templates rather than the empirical QSO2 or composite templates (and the latter solutions are excluded with 3$\sigma$ confidence). However, two of them have $\tfm$ flux densities $S(\tfm) > 300 \, \rm \mu Jy$, which strongly suggests that these galaxies could host obscured AGN \citep{cap07,des08}. The IRAC colours place one of them within the AGN region empirically determined by Stern et al.~(2005), and the other one very close to it.

The derived stellar masses of our $3<z_{\rm phot}<5$ sources range between $M \sim 10^{11}$ and  $2-3 \times 10^{12} \, \rm M_\odot$ (divide by $\sim 1.7$ for a Chabrier~(2003) IMF). The two most massive galaxies are also the two brightest $\tfm$ sources, which suggests that their rest-frame optical/near-IR light could be contaminated by an AGN component.

\subsection{Extremely dusty sources at $0.7<z_{\rm phot}<1$}

The SEDs of four $H-[4.5]>4$ galaxies have best-fitting photometric redshifts $0.7<z_{\rm phot}<1$, with very high extinction values $A_V \geq 5.0$. All of them are detected in the shallow SpUDS four-band IRAC maps, and the 5.8 and 8.0 $\rm \mu m$ photometry is crucial to decide on a best-fitting solution in these cases.  

None of these sources is detected at $\tfm$. This fact is not surprising, given that the depth of the SpUDS $\tfm$ images does not allow to detect dusty infrared normal galaxies at $z\sim1$. Although they are very obscured, the best-fitting SEDs of these sources imply that they have stellar masses $M \lsim 10^{10} \, \rm M_\odot$, so it is unlikely that they could display  large mid-infrared luminosities \citep{cap06}.

\begin{figure}
\plotone{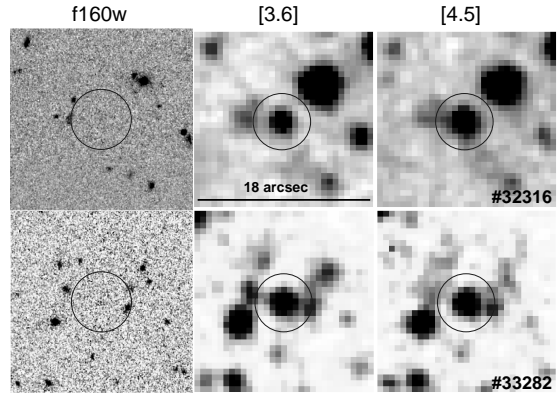}
\caption{Postage stamps of the two $z_{\rm phot}\sim 6$ galaxy candidates.\label{fig_stamps}}
\end{figure}

\subsection{$z_{\rm phot} \sim6$ galaxy candidates}

\begin{figure}
\plotone{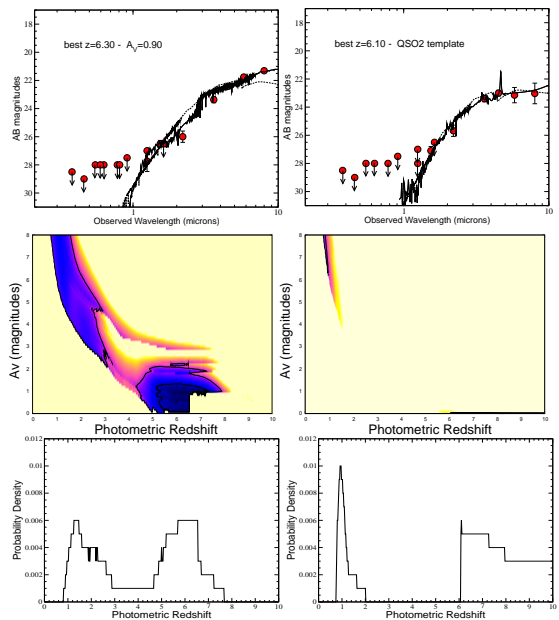}
\caption{Results of the SED fitting for \#32316 (left) and \#33282 (right). {\em Top:} the $z_{\rm phot}\sim6$ best-fitting SED (solid line) compared to the $z_{\rm phot}<2$, secondary possible SED (dashed line). {\em Middle:}  1,2 and 3$\sigma$ confidence levels on the $z_{\rm phot}-A_V$ plane, around the solution producing the minimum $\chi^2$ value. For \#33282 all the shown high-$z$ solutions appear at $A_V=0$ because they correspond to a QSO2 template for which we do not consider any extra extinction, and high-$z$ solutions with normal galaxy templates are beyond the 3$\sigma$ level, so do not appear in this plot. {\em Bottom:} Marginalised probability density $P(z_{\rm phot})$ versus $z_{\rm phot}$.   \label{fig_hizsed}}
\end{figure}

Within our $H-[4.5]>4$ sample, we find two possible $z_{\rm phot} \sim 6$ galaxy candidates (Fig. \ref{fig_stamps}). 
For source \#32316, we obtain best-fitting values $z_{\rm phot}=6.30^{+1.06}_{-1.14}$ and $A_V=0.90^{-0.70}_{+0.40}$. Source \#33282 appears to be best-fitted by the empirical QSO2 template from Polletta et al.~(2006), with a best-fitting redshift $z_{\rm phot}= 6.10^{+2.18}_{-0.10}$ (with no additional extinction). The results of our SED fitting analysis are shown in Fig.~\ref{fig_hizsed}.

For these two sources, lower $z$ solutions cannot be excluded within 2 $\sigma$ confidence, but the high-$z$ solutions are systematically preferred after re-doing the fitting on 100 random variations of each galaxy SED within the photometric error bars.  Instead,  random photometric variations applied to the lower-$z_{\rm phot}$ candidates yield that $z_{\rm phot}\gsim6$ solutions are rare (at most 10-15\% of the realisations in only six cases).  By integrating the marginalised probability density distributions $P(z_{\rm phot})$ versus $z_{\rm phot}$ (Fig.~\ref{fig_hizsed}, bottom panels) we obtain $P(z_{\rm phot}>5)=0.52$ for  \#32316, and $P(z_{\rm phot}>5)=0.75$ for  \#33282.

We note that, although these sources are formally non-detections on the ground-based near-IR images, we have manually measured the $K_s$-band fluxes at the position of their IRAC centroids, as the $K_s$-band photometry is crucial to constrain the 4000~$\rm \AA$ break in $z\sim 5-6$ sources. Also, we have independently checked the $K_s$-band photometry in the recently obtained Very Large Telescope Hawk-I images for the CANDELS/UDS field (P.I.: Fontana).

Both $z_{\rm phot}\sim6$ candidates are detected in the SpUDS images at 5.8 and 8.0 $\rm \mu m$, but neither of them is detected at  $\tfm$. We note that, if we model the SEDs of these galaxies excluding the 5.8 and 8.0 $\rm \mu m$ photometry, high-$z$ solutions are also preferred, albeit with stronger degeneracies in  $z_{\rm phot}-A_V$ space.

We investigated the use of alternative reddening laws to fit the SEDs of our $z_{\rm phot} \sim 6$ candidates. We find that the Small/Large Magellanic Clouds and  Milky Way laws produce higher $\chi^2$ values. Instead, the empirical attenuation curve proposed by Gallerani et al.~(2010) yields slightly smaller minimum $\chi^2$ values than the Calzetti et al. law, and also favours high-$z$ solutions.

The best SED fitting for \#32316 is achieved with a 0.5~Gyr-old stellar template (the age of the Universe at $z=6$ is $\sim 0.9$~Gyr).  The derived stellar mass is $\approx 3 \times 10^{11} \, \rm M_\odot$ ($\sim 1.7 \times 10^{11} \, \rm M_\odot$ with a Chabrier~(2003) IMF).  We discuss the implication of such a large stellar mass in Section \ref{sec-disc}. Nevertheless, note that it should be taken with caution, as even the observed $\ffm$ band only traces the rest-frame optical galaxy light at $z\sim6$.

\section{Near-/mid-IR colour-colour diagrams}
\label{sec-col}

\begin{figure}
\epsscale{.80}
\plotone{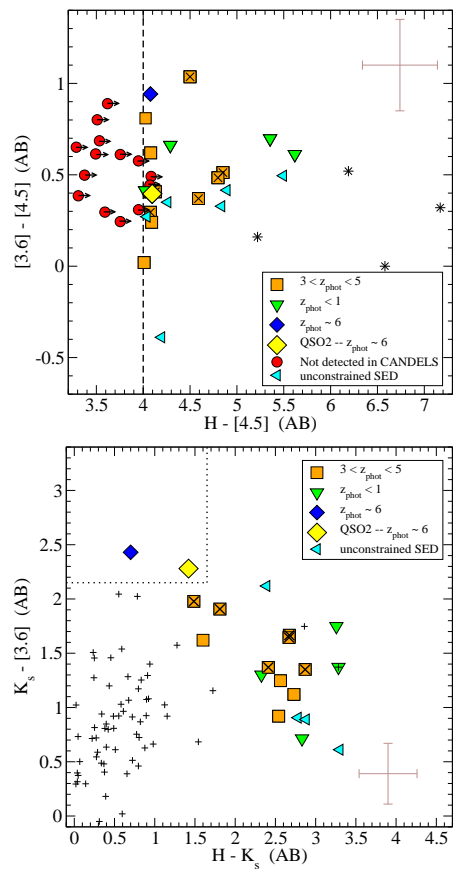}
\caption{{\em Top:}  $[3.6]-[4.5]$  versus $H-[4.5]$ colour-colour diagram for our $H-[4.5]>4$ sources. Crosses indicate sources detected at $\tfm$. Asterisks correspond to the four  $H-[3.6]>4.5$  sources analysed by Huang et al.~(2011). {\em Bottom:} $K_s-[3.6]$  versus $H-K_s$ colour-colour diagram. The dotted lines indicate the empirically-determined segregated region occupied by the $z_{\rm phot} \sim 6$ galaxy candidates. Plus-like symbols correspond to IRAC power-law AGN selected with the Donley et al.~(2012) criterion. \label{fig_colcol}}
\end{figure}

As a summary, it is very instructive to see the location of all our $H-[4.5]>4$ sources in different near-/mid-IR  colour-colour diagrams (Fig.~\ref{fig_colcol}).  Our CANDELS-detected extremely red sources display  $4<H-[4.5]<6$ colours. It is particularly noteworthy that the two $z_{\rm phot} \sim 6$ candidates are actually at the limit of our colour cut: both have $4.0<H-[4.5]<4.3$. These sources appear more unambiguously discriminated in the bottom panel of Fig.~\ref{fig_colcol}: they have $H-K_s \lsim 1.7$ and $K_s-[3.6] \gsim 2.1$.  

The $H-[4.5]>4.3$ region, instead, only contains sources whose SEDs correspond to either moderately dusty sources at $3<z_{\rm phot}<5$, or very dusty sources at $z_{\rm phot}<1$. The IRAC $[3.6]-[4.5]$ colours do not appear to discriminate between these different cases. Most of the CANDELS-detected sources with an unconstrained SED lie in a similar colour-colour regions as the  $3 < z_{\rm phot}< 5$ galaxies, so they likely are at these redshifts as well.

Three out of four of the sources analysed by Huang et al.~(2011) display  redder   $H-[4.5]$ colours than the sources analysed here. This is the consequence of the different colour selection applied by these authors ($H-[3.6]>4.5$). In turn, we do not have any source with $H-[4.5]>6$ because we do not have sufficient colour range, given the relative depths of the SEDS $\ffm$ and CANDELS $H$-band catalogues.

From the bottom panel of Fig.~\ref{fig_colcol}, we see that our $H-[4.5]>4$ sources form an orthogonal sequence in the  $K_s-[3.6]$  versus $H-K_s$  colour-colour diagram with respect to IRAC power-law AGN \citep{don12},  most of which have  bluer near-IR colours.

Although we have included our $\ffm$ sources that are $H$-band dropouts in Fig.~\ref{fig_colcol}, unfortunately the $[3.6]-[4.5]$ colours alone are not able to constrain the nature of these sources. We note, however, that their  $[3.6]-[4.5]$ colour distribution is quite similar to that of our CANDELS-detected sources, so we expect a similar variety in their nature.

\section{Discussion}
\label{sec-disc}

The CANDELS-detected $H-[4.5]>4$ galaxy population analysed here appears to be composed of different kinds of sources, and their extremely red near-to-mid-IR colours can be produced by different combinations  of redshifts and internal extinction values.

At least 45\% of our CANDELS-detected, $H-[4.5]>4$ sources are dust-obscured, massive galaxies at $3<z_{\rm phot}<5$, similar in nature to the sub-millimetre source GN10. This percentage would be as high as 75\% if the eight sources with an unconstrained SED were also at $3<z_{\rm phot}<5$. Two of them are powerful $\tfm$ sources, which indicates that they likely host an obscured AGN. The four sources analysed by Huang et al.~(2011) have similar properties. Rodighiero et al.~(2007) also studied similar galaxies within their IRAC-selected sample with faint near-IR counterparts, although formally the vast majority of their sources display bluer  $H-[4.5]$ colours than those characterising our sample.

Within our  $H-[4.5]>4$ galaxy population we have also found two possible $z_{\rm phot}\sim6$ candidates. However, alternative low-$z$ solutions cannot be discarded within $2\sigma$ confidence, so we cannot confirm the high-$z$ nature of these sources. In spite of this, we note that our two sources are more likely true $z_{\rm phot}\sim6$ galaxies than the IRAC high-$z$ candidates analysed by other authors \citep{mob05,wik08}, as our objects are considerably fainter in the IRAC bands and are not detected in the SpUDS $\tfm$ images. 

For the $z_{\rm phot}\sim6$ candidate that is  best-fitted with a stellar template, we estimate a stellar mass of  $\sim 3 \times 10^{11} \, M_\odot$ ($\sim 1.7 \times 10^{11} \, M_\odot$ with a Chabrier~(2003) IMF). This value is significantly higher than the stellar masses derived for more typical galaxies at such high redshifts. 

The presence of massive galaxies at $z\sim6$ in a small cosmological volume is difficult to reconcile with the predicted comoving number density of dark-matter haloes that would be able to host such galaxies. According to the Sheth \& Tormen~(1999) formalism, the number density of $M> 10^{13} \, \rm M_\odot$ haloes at $z\sim6$ is only $10^{-9} \, \rm Mpc^{-3}$, three orders of magnitude smaller than the density of our  two  $z_{\rm phot}\sim6$ candidates over the SEDS/CANDELS field.  However,  the predicted number density of haloes with $M> 10^{12.5} \, \rm M_\odot$ is $10^{-7} \, \rm Mpc^{-3}$, which means that different uncertainties in e.g. the derived stellar masses, the IMF, and/or the assumed baryon conversion fraction could make that these discrepancies are actually less dramatic. 

We note that such large stellar masses have been suggested for confirmed $z\sim6$ quasars. Particularly, Valiante et al.~(2011) argued that the observed dust mass and metallicity of SDSS J1148+5251 ($z=6.4$) can only be reproduced with a standard IMF if the stellar mass is of $\sim 1-5 \times 10^{11} \, \rm M_\odot$. Instead, reconciling the chemical properties of this quasar with smaller stellar masses requires a top-heavy IMF. Although SDSS quasars are much rarer than our two sources, this argument suggests that one cannot reject high-$z$ candidates only on the ground of large derived stellar masses.

Recently, based on the analysis of the 1.5 mag shallower SpUDS IRAC data, Caputi et al.~(2011) studied the presence of massive galaxies at high redshifts and found only a few candidates at $z_{\rm phot}>5$.  Our finding of two possible galaxy candidates at $z_{\rm phot}\sim 6$ in the SEDS IRAC maps would suggest that deeper mid-IR data is essential to properly constrain the presence of massive galaxies at $z>5$. The two  $z_{\rm phot}\sim6$ candidates found here are detected in the SpUDS images, but were among the $\sim$8\% of faint ($[4.5]>22.4$ mag) SpUDS IRAC sources not analysed by Caputi et al.~(2011) due to the lack of any near-IR identification at that time.

Our $z_{\rm phot}\sim 6$ candidates are so faint at near-IR magnitudes that their spectroscopic confirmation is beyond the capabilities of any existing near-IR spectrograph. Instead, the internal extinctions of these galaxies suggest that they could be far-IR sources, and thus the Atacama Large Millimetre Array (ALMA) could currently offer the only possibility to confirm their high $z$.

Overall, we conclude that our red galaxy sample shows the tip of the iceberg of a previously unrecognised $z>3$ galaxy population, which will only be fully revealed with the advent of the Mid-Infrared Instrument (MIRI) on the future {\em James Webb Space Telescope} \citep{wri04}.

\begin{deluxetable}{lccccccccl}
\tabletypesize{\scriptsize}
\tablecaption{Coordinates, photometry and best $z_{\rm phot}$ estimates for the $H-[4.5]>4$ galaxies.  \label{tbl}}
\tablewidth{0pt}
\tablehead{\colhead{ID} & \colhead{RA (J2000)} & \colhead{DEC(J2000)} & \colhead{F814W} &
\colhead{F125W} & \colhead{F160W} &  \colhead{$K_s$} & \colhead{[3.6]} & \colhead{[4.5]} & \colhead{$z_{\rm phot}$}}
\startdata
\#32316 &  02:17:43.32  & -05:11:57.4 & $>28.0$ & $27.74\pm0.74$ & $26.50\pm0.24$ & 
$25.80 \pm0.40$ &  $23.37\pm0.22$ & $22.43\pm0.15$ & $6.30^{+1.06}_{-1.14}$ \\
\#33282 &  02:18:05.80  & -05:11:23.1 & ---\tablenotemark{a} & $>27.5$ & $27.10\pm0.30$ & 
$25.60 \pm 0.50$ & $23.40\pm0.22$ & $23.00\pm0.20$ & $6.10^{+2.18}_{-0.10}$ \\
\#27564 &  02:17:16.35 &  -05:14:43.1 & $26.60\pm0.30$ & $25.00\pm0.05$ & $24.89\pm0.05$ &
$23.40\pm0.06$ & $21.42\pm0.10$ & $20.39\pm0.10$ & $4.94^{+0.56}_{-0.18}$ \\
\#26857 & 02:17:51.69 & -05:15:07.2 & $26.95\pm0.64$ & $25.49\pm0.16$ & $24.39\pm0.14$ &
 $22.58\pm0.07$ & $20.67\pm0.10$ & $20.26\pm0.10$ &  $4.92^{+0.46}_{-0.14}$\\
 \#36664 & 02:17:32.40 & -05:09:23.0 & $28.44\pm0.95$ & $27.85\pm0.72$ & $27.72\pm0.68$ &
$24.85\pm0.35$ & $23.50\pm0.24$ & $23.13\pm0.22$ & $4.88^{+1.32}_{-1.98}$ \\
\#25692 & 02:17:03.45 & -05:15:51.3 & $27.50\pm0.92$ & $27.91\pm0.72$ & $26.82\pm0.33$ &
$25.22\pm0.40$ & $23.60\pm0.23$ & $22.79\pm0.15$ & $4.86^{+2.38}_{-1.32}$ \\
\#35497 & 02:18:14.99 & -05:10:02.7 & $>28.0$ & $27.35\pm 0.55$ & $>27.5$ & 
$25.88 \pm 0.53$ & $24.05\pm0.25$ & $23.86\pm0.22$ & $4.40^{+1.24}_{-1.48}$\tablenotemark{b}\\
\#34284 & 02:18:05.91 & -05:10:49.0 & ---\tablenotemark{a} & $26.63\pm0.21$ & $26.88\pm0.27$ &
$24.15\pm0.12$ & $23.03\pm0.17$ & $22.79\pm0.18$ & $4.22^{+0.84}_{-0.64}$ \\
\#25215 & 02:17:36.95 & -05:16:07.2 & $>28.0$ & $>27.5$ & $27.35\pm0.53$ & 
$24.81\pm0.20$ & $23.89\pm0.24$ & $23.27\pm0.20$ & $4.10^{+0.70}_{-0.92}$ \\
\#26909 & 02:17:20.77 & -05:15:06.9 & $27.80\pm0.93$ & $>27.5$ & $26.64\pm0.28$ & 
$23.97\pm0.10$ & $22.30\pm0.16$ & $21.79\pm0.12$ & $4.04^{+0.78}_{-0.54}$ \\
\#38455 & 02:17:22.96 & -05:08:15.7 & $>28.0$ & $26.66\pm0.31$ & $27.27\pm0.46$ &
$24.71\pm0.20$ & $23.46\pm0.31$ & $23.44\pm0.27$  & $3.74^{+2.18}_{-0.44}$ \\
\#30425 & 02:17:40.51 & -05:13:10.6  & $27.26\pm0.84$ & $26.25\pm0.18$ & $26.05\pm0.15$ &
$23.38\pm0.07$ & $21.74\pm0.16$ & $21.26\pm0.12$ & $3.52^{+1.42}_{-0.24}$\\
\#37742 & 02:17:13.13 & -05:08:43.4 & $26.36\pm0.19$ & $25.69\pm0.13$ & $26.20\pm0.18$ &
$23.78\pm0.09$ & $22.41\pm0.15$ & $22.12\pm0.15$ & $3.42^{+0.34}_{-0.10}$ \\
\#34174 & 02:18:05.65 & -05:10:49.7 & ---\tablenotemark{a} & $27.05\pm0.35$ & $27.92\pm1.21$ &
 $24.66\pm0.18$ & $22.91\pm0.16$ & $22.30\pm0.14$ & $1.00^{+0.64}_{-0.04}$\\
\#38833 & 02:17:22.43 & -05:08:05.1 & $27.00\pm0.55$ & $>27.5$ & $26.79\pm0.29$ &
 $24.46\pm0.16$ & $23.16\pm0.26$ & $22.50\pm0.21$  & $0.98^{+0.32}_{-0.44}$ \\
\#34264 & 02:18:05.04 & -05:10:46.3 & ---\tablenotemark{a} & $26.36\pm0.29$ & $27.68\pm0.97$ &
$24.40\pm0.14$ & $23.03\pm0.16$ & $22.33\pm0.13$ & $0.90^{+0.06}_{-0.04}$ \\
\#32715 & 02:18:03.69 & -05:11:42.5 & ---\tablenotemark{a} & $28.82\pm1.67$ & $27.52\pm0.84$ &
$24.69\pm0.19$ & $23.98\pm0.23$ & $23.56\pm0.21$ & $0.72^{+0.36}_{-0.10}$ \\
\#26375 & 02:17:00.12 & -05:15:24.7 & $27.06\pm0.56$ & $>27.5$ & $27.81\pm0.75$ &
$24.93\pm0.21$ & $24.04\pm0.21$ & $23.77\pm0.22$ & ?\\
\#26721 & 02:17:09.69 & -05:15:11.1 & $>28.0$ & $>27.5$ & $27.77\pm0.78$ &
$>26.5$ & $24.29\pm0.22$ & $23.68\pm0.19$ & ? \\ 
\#28839 & 02:18:11.52 & -05:14:01.6 &  ---\tablenotemark{a} & $26.44\pm0.31$ & $>27.5$ &
$24.86\pm0.20$ & $23.53\pm0.22$ & $23.04\pm0.17$ & ?\\
\#29434 & 02:18:03.58 & -05:13:34.6 & ---\tablenotemark{a} & $26.65\pm0.72$ & $27.75\pm1.15$ &
$24.45\pm0.15$ & $23.84\pm0.27$ & $23.49\pm0.23$ & ? \\
\#29731 & 02:17:21.78 & -05:13:26.5 & $>28.0$ & $25.94\pm0.37$ & $26.99\pm0.70$ &
$24.21\pm0.12$ & $23.30\pm0.19$ & $22.89\pm0.17$ & ?\\
\#30561 & 02:17:03.44 & -05:12:56.0 & $27.69\pm1.13$ & $28.64\pm 1.76$ & $26.95\pm0.38$ & 
$>26.5$ & $22.37\pm0.16$ & $22.76\pm0.18$ & ? \\
\#33716 & 02:17:57.06 & -05:11:04.9 & $27.46\pm1.72$ & $26.32\pm0.19$ & $26.89\pm0.33$  &
$24.50\pm0.45$ & $22.38\pm0.18$ & $22.05\pm0.17$ & ?\\
\#35848 & 02:17:06.26 & -05:09:48.3 & $>28.0$ & $27.68\pm0.85$ & $>27.5$ &
$>26.5$ & $24.18\pm0.36$  & $23.50\pm0.28$  & ? 
\enddata
\tablenotetext{a}{Out of ACS coverage field.}
\tablenotetext{b}{A solution $z_{\rm phot}=0.74^{+0.46}_{-0.14}$ also appears within 1$\sigma$ confidence.}
\end{deluxetable}

\end{document}